# RAPID OSCILLATIONS IN CATACLYSMIC VARIABLES.

# XV. HT CAMELOPARDALIS (= RX J0757.0+6306)

Jonathan Kemp,[1,2] Joseph Patterson,[2] John R. Thorstensen,[3]

Robert E. Fried,[4] David R. Skillman,[5] and Gary Billings[6]

[1] Joint Astronomy Centre, University Park, 660 North A`ohōkū Place, Hilo, HI 96720; j.kemp@jach.hawaii.edu

[2] Department of Astronomy, Columbia University, 550 West 120th Street, New York, NY 10027; jop@astro.columbia.edu

[3] Department of Physics and Astronomy, Dartmouth College, 6127 Wilder Laboratory, Hanover, NH 03755; thorstensen@dartmouth.edu

[4] Center for Backyard Astrophysics (Flagstaff), Braeside Observatory, Post Office Box 906, Flagstaff, AZ 86002; captain@asu.edu

[5] Center for Backyard Astrophysics (East), 9517 Washington Avenue, Laurel, MD 20723; dskillman@home.com

[6] Center for Backyard Astrophysics (Calgary), 2320 Cherokee Drive NW, Calgary, AB T2L 0X7, Canada; obs681@telusplanet.net

**ABSTRACT**

We present photometry and spectroscopy of HT Camelopardalis, a recently discovered X-ray-bright cataclysmic variable. The spectrum shows bright lines of H, He I, and He II, all moving with a period of 0.059712(1) d, which we interpret as the orbital period. The star's brightness varies with a strict period of 515.0592(2) s, and a mean full amplitude of 0.11 mag. These properties qualify it as a *bona fide* DQ Herculis star (intermediate polar) — in which the magnetism of the rapidly rotating white dwarf channels accretion flow to the surface. Normally at $V$=17.8, the star shows rare and very brief outbursts to $V$=12–13. We observed one in December 2001, and found that the 515 s pulse amplitude had increased by a factor of ~100 (in flux units). A transient orbital signal may also have appeared.

*Subject headings*: accretion, accretion disks — binaries: close — novae, cataclysmic variables — stars: individual (HT Cam)

# 1. INTRODUCTION

In this series of papers, we study periodic signatures of cataclysmic variables (CVs) on timescales shorter than the binary orbital period. CVs show many types of variability, but our understanding of the up-and-down excursions in light is usually pretty meager, except for the rare case where the variability is strictly periodic. This phenomenon was discovered in the old nova DQ Herculis by Merle Walker (Walker 1954, 1956), then mostly forgotten until the discovery of pulsars reawakened interest in rapid periodic signals. The 1970s saw more work on this phenomenon, and led to a working model: a magnetic and rapidly rotating white dwarf, channeling the flow of accreting gas onto the star's magnetic poles (Lamb 1974, Katz 1975, Patterson 1979). The model has passed the test of time, with only a few blemishes not yet healed or understood. These binaries are now called DQ Herculis stars, or *intermediate polars*; about 20–25 are reasonably well-credentialed and discussed in recent reviews (Patterson 1994, Hellier 1998).

Because accreting gas falls more or less directly to the white dwarf surface, a strong shock develops above the surface and is heated to $\sim 10^8$ K, radiating hard X-rays. Thus these stars tend to be among the stronger X-ray sources seen in CVs, and most of them are now first identified from X-ray surveys. From the ROSAT All-Sky Survey, Tovmassian et al. (1998, hereafter T98) identified RX J0757.0+6306 as an 18th magnitude blue star with emission lines and suggested periods of 81 and 8.5 minutes (candidates for $P_{orb}$ and $P_{spin}$, respectively). The star was later renamed HT Camelopardalis. We began an intensive observing campaign and here report the results, basically verifying the periodic signals and providing long-baseline ephemerides.

# 2. SPECTROSCOPY

We obtained spectra of HT Cam in 1997 December, 1998 January, and 1998 March using the Hiltner 2.4 m telescope at MDM Observatory, the Modular spectrograph, and a thinned $2048^2$ Tektronix CCD yielding 2 Å/pixel from 4000 to 7500 Å. The equipment and procedures were essentially as described by Thorstensen, Taylor, & Kemp (1998). The signal-to-noise ratio decreased towards the ends because of vignetting in the spectrograph camera. The 1 arcsec slit gave a spectral resolution of ~3.5 Å, and frequent exposures of Hg, Ne, and Xe comparison lamps were fitted with root-mean-square residuals ~0.06 Å. We observed flux standards and hot stars to allow conversion to absolute flux, and reduced the raw pictures to flux vs. wavelength using standard IRAF procedures.[1] Our time baseline was 98 days and the spectra cover a 7.2 h range of hour angle. Individual exposures were 6 min in 1997 December, 10 min in 1998 January, and 7 min in 1998 March.

The average spectrum (Figure 1) appears typical for dwarf novae at minimum light, though with rather strong emission lines, detailed in Table 1. Hα has a half width at half maximum of 700 km/s, and the line can be traced out to ±1400 km/s from line center. The lines are slightly double-peaked, with Hα showing two peaks separated by 500 km/s. HeII λ4686 is quite noticeable but not remarkably strong. The continuum is moderately blue, and at this signal-to-noise ratio no contribution from an M-dwarf secondary is detected. The flux level, which we estimate to be uncertain by ~30 percent, implies $V$~17.8.

---

[1] IRAF, the Image Reduction and Analysis Facility, is distributed by the National Optical Astronomy Observatories.

We measured velocities of Hα by convolving the line profile with the derivative of a Gaussian and looking for the zero of the convolution (Schneider & Young 1981). A "residual-gram" period search (Thorstensen et al. 1996) shows a strong signal near 16.7 cycle/d (Figure 2). Because of the long observing sessions in 1997 December, the daily cycle count is secure. The cycle count between observing sessions almost as well determined, with the Monte Carlo test of Thorstensen & Freed (1985), giving a discriminatory power near 997/1000; much less likely choices are separated from the best frequency by ~1/48 cycle/d (Figure 2, top panel). The best-fitting sinusoid of the form

$$\upsilon(t) = \gamma + K \sin [2\pi (t - T_0) / P]$$

has

$$\begin{aligned} T_0 &= \text{HJD } 2450842.6636 \pm 0.0007, \\ P &= 0.059712 \pm 0.000001 \text{ d}, \\ K &= 69 \pm 4 \text{ km/s}, \\ \gamma &= -69 \pm 3 \text{ km/s, and} \\ \sigma &= 23 \text{ km/s}, \end{aligned}$$

where $\sigma$ is the scatter around the best fit. Figure 3 shows the velocities folded on the best ephemeris. The period we derive, 85.985 min, is consistent with the longest of the three possible daily aliases given by T98, but their preferred 81-min period is excluded. Although we believe that $K$ is not a reliable indicator of the motion of either star, the radial-velocity period is almost certainly $P_{orb}$.

## 3. LIGHT CURVES AND PERIODS

We obtained time-series differential photometry at three stations of the Center for Backyard Astrophysics network (Braeside Observatory, CBA-Calgary and CBA-East; Skillman 1993), and the 1.3 m and 2.4 m telescopes of MDM Observatory. Covering 46 nights during 1997–2002, the observations were strategized to enable a precise period estimate during each year, with some long and dense time series (10 hrs/night) to define the period structure accurately. In order to maximize signal-to-noise, we made the bulk of the observations in unfiltered (4000–8000 Å) light. A few nights used broadband filters ($B$, $V$, $I$); the periodic signatures appeared to be the same in all filters, so we did not distinguish them in subsequent analysis.

During most of the observations HT Cam remained within 0.4 mag of $V$=17.8, its usual quiescent level; but we did observe one brief outburst to $V$~12.5 in December 2001.

### 3.1 LIGHT CURVE AND PERIODS

On nearly every night, the light curve looked essentially identical to the upper frame of Figure 4: erratic ripples of 0.1–0.2 mag superimposed on the steady beat of an 8.5 minute clock. The full (peak-to-trough) amplitude of the latter signal varied from 0.06 to 0.15 mag from night to night. We calculated power spectra for each night of adequate length, and for several multi-night segments. Essentially every night shows a signal at 168 c/day, with no other credible signals; in particular, the full-amplitude upper limit for any periodic signal near $P_{orb}$ is 0.05 mag.

The lower frame of Figure 4 contains part of a 17-night power spectrum, with the obvious strong signal marked. The 515.06±0.01 s signal is present with a full amplitude of 0.114±0.008 mag, and the surrounding structure in the power spectrum is entirely due to the spectral "window" (disappears entirely when a pure signal of that period and amplitude is subtracted). No other signal at frequencies >100 c/day exceeds 0.022 mag in full amplitude (except for a weak first harmonic at 335.50 c/day). Inset in the lower frame is the mean waveform of the 515 s signal, a nearly pure sinusoid.

It was easy to measure timings of 515 s maxima by fitting sinusoids to the light curve, for all nightly observations exceeding ~1 hour in length. The resultant pulse timings are listed in Table 3.

### 3.2 LONG–TERM PULSE EPHEMERIS

All the seasonal periods are consistent with a period of 515.06±0.02 s. This stability enabled secure cycle counting during the ~200 days between observing seasons, and we found that essentially all the timings can be satisfied with a (unique) period of 515.0592±0.0002 s. In Figure 5 we show all the timings reduced to an O–C diagram with this test period. The fitted line corresponds to

Pulse maximum = HJD 2,450,798.7443(2) + 0.005961333(3) $E$,

with $|\dot{P}| < 2\times10^{-11}$.

### 3.3 HT CAM IN OUTBURST

Like most short-period CVs, HT Cam shows outbursts, which are at least roughly like dwarf-nova eruptions. The VSNET archive lists three certain eruptions in a 1.6 year period. But the scarcity of observation and very short lifetime of the outbursts imply that some are missed, suggesting a recurrence time in the range ~100–150 days.

We managed to secure time-series photometry of the December 2001 outburst. On JD 270.0 and 271.9, we obtained time series of the star at quiescence, with the pulse phase and amplitude typical of quiescence. Over the interval JD 273.18–273.33, Hitoshi Itoh reported the star in full outburst at $V$=12.2 (reported by T. Kato in vsnet-alert 6944). We obtained long coverage later that night (273.63–274.11), which showed the star in rapid decline (~0.2 mag/day). This is shown in the upper frame of Figure 6.

The middle frame of Figure 6 shows a power spectrum of that 12-hour observation, with detectable signals near the known orbital and pulse frequencies. The lowest frame of Figure 6 shows a portion of that night's light curve. Tick marks, spaced at 515.06 s intervals, illustrate the strong pulse in this segment. The period during the observation agrees with the known pulse period to within measurement error (~1%). But the mean pulse maximum arrived at HJD 2452273.7419, displaced by +0.46±0.05 cycles on the above ephemeris; and the pulse amplitude in flux units was increased over that of quiescence, by a factor of ~100.

We analyzed 8 independent segments of that night's data in this manner, and found similar results: the pulse amplitude was always increased over that of quiescence by a factor of 20-100, and the pulse maximum was delayed by 0.22–0.51 (typically ±0.07) cycles. As can be guessed from the fatness of the power-spectrum peak, the signal is not quite coherent, though most of the power remains at the familiar period.

On subsequent nights, JD 274.7 and 276.7, the star had dropped to $V$=16.3, with a pulse amplitude and phase characteristic of quiescence.

## 4. DISCUSSION

### *4.1 HT CAM SHOWS ITS COLORS: A DQ HER STAR*

A period of this stability ($Q=|\dot{P}|>10^{10}$) is not commonly found in stars, and never in accretion disks where the intrinsic shear prevents any periodic signal from achieving high stability. Of course, binary orbital periods are very stable, but we are confident that the radial velocities correctly identify $P_{orb}$ as 0.059712 d. The other natural candidate for a very stable period is rigid-body rotation of the white dwarf. Rotation, and in particular the rapid rotation of a magnetic white dwarf, is the clock that underlies the optical and X-ray signals which are the principal defining signature of a DQ Herculis star (also called intermediate polar). About two dozen such stars are now known, and the demonstrable high stability of the 515 s clock basically certifies HT Cam as a new member of the class.

These stars also have an unusually high $L_x/L_{opt}$ ratio among CVs, presumably because radially channeled gas is likely to surrender its infall energy in a strong shock at the surface. HT Cam has $F_x$(0.1–2.8 keV) / $F_{opt}$(5000–6000 Å) = 4, one of the highest values seen in CVs (cf. Richman 1996). So this is extra supporting evidence. It is nearly certain that a sensitive X-ray time series of HT Cam will show strong hard X-ray pulsations at 515 s.

### *4.2 MAGNETISM*

At first thought, one would expect all DQ Hers to spin up rapidly due to the accretion of angular momentum from the disk. But unless they all began accreting the day before yesterday, they are rotating too slowly to be consistent with this simple prescription. It is more likely that they have reached "spin equilibrium", where the spin-up torques from accreting matter balance the spin-down torques from ejected matter and the dragging of field lines in the outer disk. The theory for this equilibrium was well explained by Ghosh & Lamb (1978), and lucidly applied by Norton & Watson (1989). A magnetic white dwarf in spin equilibrium with $P$=515 s and $\dot{M}$ ~$10^{16}$ g/s (appropriate for this $P_{orb}$) should have a magnetic moment ~$10^{33}$ G-cm$^3$.

### *4.3 OUTBURSTS*

The outbursts of HT Cam are a peculiar feature. So far, no evidence has emerged which reveals their origin, and it is premature to assume that they are dwarf-nova outbursts (viz. arising from thermal instability in the disk). It is reasonably clear, though, that the rate of mass flow to the white dwarf — the "accretion rate" — is greatly increased during outburst, since the 515 s pulse amplitude in flux units should track that pretty well as long as the field lines are rigid enough to steer accretion flow.

Some additional evidence on the nature of the outburst might be present in the data on the rapid periodic signal (signals near the spin period). The greatly increased flux in the 515 s signal testifies pretty clearly to the higher accretion rate. The existence of a temporary large phase shift (+0.46 cycles in the example cited; +0.38 cycles averaged over the night) is not problematic, since the timing of maximum light is sensitive to the exact geometrical details of the accretion flow. (In particular, a tenfold increase in $\dot{M}$ could through associated optical-depth effects plausibly make a 0.5 cycle change in the front/back asymmetry which is at the heart of the observed signal — and, for that matter, might even switch the identity of the accreting pole.) Unfortunately we are not yet able to sort among these possibilities and find a best interpretation of the observed phenomena.

Additional discussion of outbursts in DQ Her stars has been given by Szkody & Mateo (1984), Hellier et al. (2000), Chapter 7.7 of Warner (1995), and Chapter 9 of Hellier (2001). In particular, the outbursts of EX Hya, studied by Hellier et al. (2000), appear to be a pretty good match for those of HT Cam.

## 5. SUMMARY

1. Our observations confirm the suggestion of T98 that HT Cam is a DQ Herculis star (intermediate polar) with $P_{spin}$=515 s. The precise value of the spin period is 515.0592(2) s — or possibly an orbital sideband. X-ray observations should reveal strong X-ray pulses at this period.

2. The emission-line spectrum is unremarkable for a low-$\dot{M}$ CV. The lines move with a period of 85.9853(14) minutes, which we interpret as the underlying orbital period — the shortest $P_{orb}$ established among the well-certified DQ Her stars.

3. Outbursts are rare and very brief; the December 2001 outburst was above $V$=15.5 mag for only 2 days. The spin pulse was strengthened by a factor of ~100 in outburst, as has been observed in several other DQ Her stars (EX Hya, GK Per, XY Ari). The signal appears to have shown also a transient phase shift of +0.3–0.5 cycles.

4. We do not yet have enough information to classify these as dwarf-nova outbursts; but if they are, then superoutbursts are expected from a dwarf nova of this $P_{orb}$. A superoutburst in a certifiably magnetic CV (a DQ Her star) would provide a grand experiment in the interplay of magnetism and accretion flow.

This research was supported in part by NSF grants 9987334 and 0098254, and grant GG–0042 from the Research Corporation.

TABLE 1
Hα RADIAL VELOCITIES

| HJD[a] | V[b] | HJD[a] | V[b] | HJD[a] | V[b] | HJD[a] | V[b] |
|---|---|---|---|---|---|---|---|
| 800.7765 | −72 | 800.9686 | −148 | 801.8946 | 43 | 893.6963 | −113 |
| 800.7814 | −119 | 800.9755 | −114 | 801.8993 | 47 | 893.7093 | −133 |
| 800.7861 | −126 | 800.9802 | −53 | 801.9041 | −35 | 893.7148 | −65 |
| 800.7909 | −127 | 800.9850 | −66 | 801.9643 | −35 | 893.7202 | −59 |
| 800.7956 | −159 | 800.9897 | −22 | 801.9690 | −71 | 893.7257 | −10 |
| 800.8004 | −103 | 800.9945 | 4 | 801.9738 | −104 | 893.7312 | 11 |
| 800.8051 | −47 | 800.9992 | 26 | 801.9785 | −113 | 893.7366 | −27 |
| 800.8099 | −25 | 801.0040 | −12 | 801.9833 | −103 | 897.7127 | −110 |
| 800.8166 | −4 | 801.0087 | −40 | 801.9880 | −113 | 897.7182 | −43 |
| 800.8214 | −17 | 801.0184 | −122 | 842.6463 | −125 | 897.7236 | −21 |
| 800.8261 | −2 | 801.0232 | −146 | 842.6538 | −163 | 897.7291 | −1 |
| 800.8309 | −29 | 801.0279 | −159 | 842.6613 | −117 | 897.7345 | 3 |
| 800.8356 | −82 | 801.0327 | −194 | 842.6689 | −112 | 897.7400 | −23 |
| 800.8404 | −101 | 801.0374 | −141 | 842.6764 | −11 | 897.7477 | −69 |
| 800.8451 | −103 | 801.0422 | −129 | 842.6900 | −66 | 897.7531 | −110 |
| 800.8499 | −122 | 801.0469 | −52 | 842.6975 | −106 | 897.7586 | −145 |
| 800.8571 | −91 | 801.0517 | −71 | 842.7050 | −121 | 897.7640 | −146 |
| 800.8618 | −50 | 801.0587 | −13 | 842.7125 | −160 | 897.7695 | −163 |
| 800.8666 | −52 | 801.7974 | −60 | 845.8022 | −97 | 897.7749 | −88 |
| 800.8713 | −13 | 801.8024 | −110 | 845.8097 | −138 | 897.7831 | −24 |
| 800.8761 | 9 | 801.8072 | −139 | 845.8172 | −123 | 897.7886 | 1 |
| 800.8808 | 16 | 801.8119 | −112 | 845.8248 | −103 | 897.7940 | 28 |
| 800.9353 | −30 | 801.8167 | −65 | 845.8323 | −27 | 898.6888 | 0 |
| 800.9401 | 27 | 801.8214 | −67 | 845.8458 | 9 | 898.6942 | −14 |
| 800.9448 | −37 | 801.8262 | −53 | 893.6690 | −30 | 898.7014 | −40 |
| 800.9496 | −41 | 801.8309 | −32 | 893.6744 | −19 | 898.7068 | −63 |
| 800.9543 | −84 | 801.8384 | −29 | 893.6799 | −20 | 898.7123 | −119 |
| 800.9591 | −86 | 801.8432 | 9 | 893.6853 | −91 | 898.7177 | −130 |
| 800.9638 | −124 | 801.8574 | −107 | 893.6908 | −120 | | |

[a] Heliocentric JD of mid-integration minus 2,450,000.
[b] Heliocentric Hα velocity (km/s).

TABLE 2
EMISSION LINES

| Feature | E.W.[a] (Å) | Flux ($10^{-14}$ erg/cm$^2$/s) |
|---|---|---|
| H$\delta$ | 44: | 2.3: |
| H$\gamma$ | 77 | 3.4 |
| HeI $\lambda$4471 | 23 | 0.9 |
| HeII $\lambda$4686 | 14 | 0.5 |
| H$\beta$ | 121 | 4.4 |
| HeI $\lambda$4921 | 9: | 0.3: |
| HeI $\lambda$5015 | 8 | 0.30 |
| FeII $\lambda$5169 | 7: | 0.2: |
| HeI $\lambda$5876 | 37 | 1.0 |
| H$\alpha$ | 161 | 4.1 |
| HeI $\lambda$6678 | 16 | 0.40 |
| HeI $\lambda$7607 | 12: | 0.25: |

[a] Emission is counted as positive.

TABLE 3
TIMES OF 515 s PULSE MAXIMA

| HJD 2,450,000+ | | | |
|---|---|---|---|
| 798.7435 | 855.5978 | 1293.6239 | 2245.9175 |
| 799.6800 | 856.6110 | 1296.6225 | 2246.8715 |
| 808.7950 | 858.6737 | 1327.6403 | 2250.8595 |
| 810.5837 | 896.7607 | 1330.6391 | 2271.7779 |
| 813.9042 | 908.6775 | 1623.6018 | 2273.6638 |
| 841.6064 | 1218.5892 | 1682.6791 | 2274.6270 |
| 842.5781 | 1264.7775 | 1687.6391 | 2276.6242 |
| 843.5795 | 1265.6533 | 1899.6714 | |
| 850.8820 | 1285.6663 | 1932.5902 | |
| 851.6331 | 1287.6452 | 2162.9899 | |

# FIGURE CAPTIONS

FIGURE 1. — Mean spectrum of HT Cam. The flux scale is uncertain by perhaps 50 per cent because of variable clouds and seeing.

FIGURE 2. — Period searches of the H$\alpha$ velocities. The *lower panel* shows the local maxima in the period search connected by straight lines, while the *upper panel* gives a detailed view in the vicinity of the highest peak.

FIGURE 3. — Radio velocities of the H$\alpha$ line folded on the best ephemeris. Data are from 1997 December (squares), 1998 January (triangles), and 1998 March (circles). The best fit sinusoid is superposed. All data are plotted twice for continuity.

FIGURE 4. — *Upper frame*, sample light curve of HT Cam in quiescence, showing erratic flickering and a 515 s periodic signal. *Lower frame*, a portion of the power spectrum of a 17-night light curve; the only significant signal marked with its frequency in cycles/day. The full amplitude of that signal is 0.114 mag, and the mean waveform (inset) is closely sinusoidal. All the additional structure around 167 c/day arises merely from the spectral window of the time series.

FIGURE 5. — O–C diagram of the timings of maximum light in the 515 s signal, relative to the test ephemeris: HJD 2450798.7435 + 0.005961333 *E*.

FIGURE 6. — HT Cam in outburst, 30 December 2001. *Upper frame*, 12-hour light curve, showing decline at ~3 mag/day. Middle frame, power spectrum, showing signals at the known orbital and pulse frequencies (within the ±0.4 c/day error). *Lower frame*, close-up view of a portion of the light curve (at *V*=13.2), with the inset tick marks spaced at 515 s. The train of 10 marked pulses shows a mean period of 511±7 s, and a mean amplitude of 0.18 mag. Pulse trains like this appeared sporadically through the night.

Fig 1

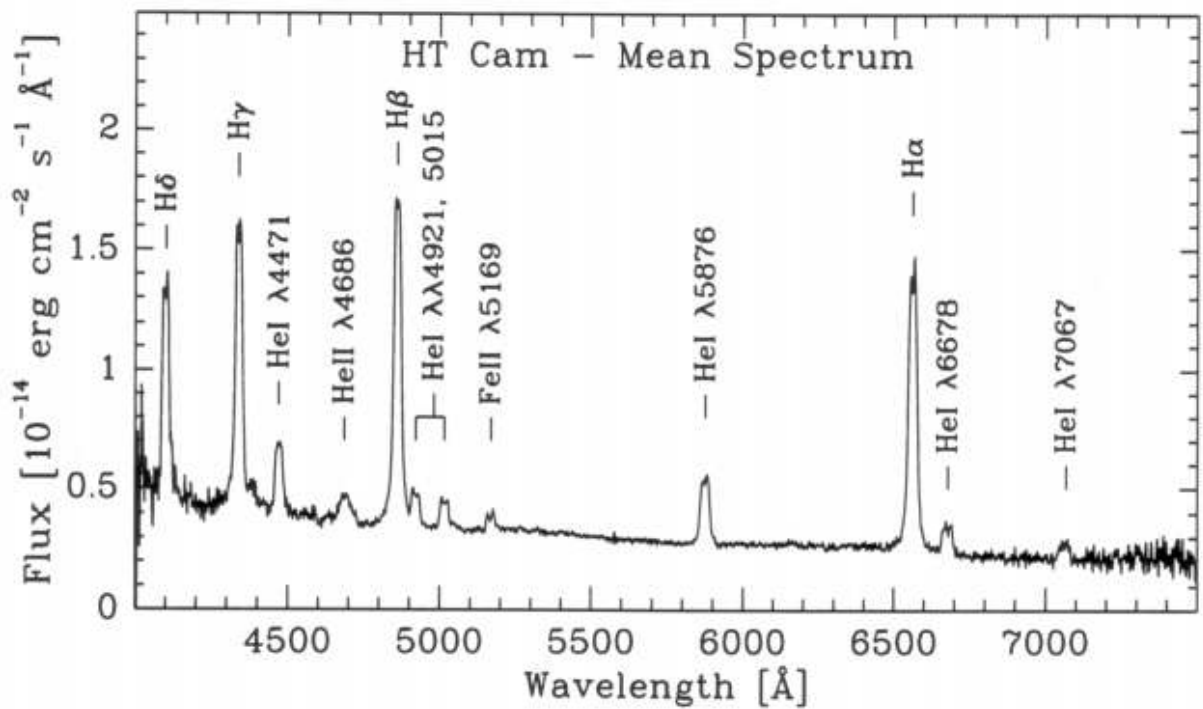

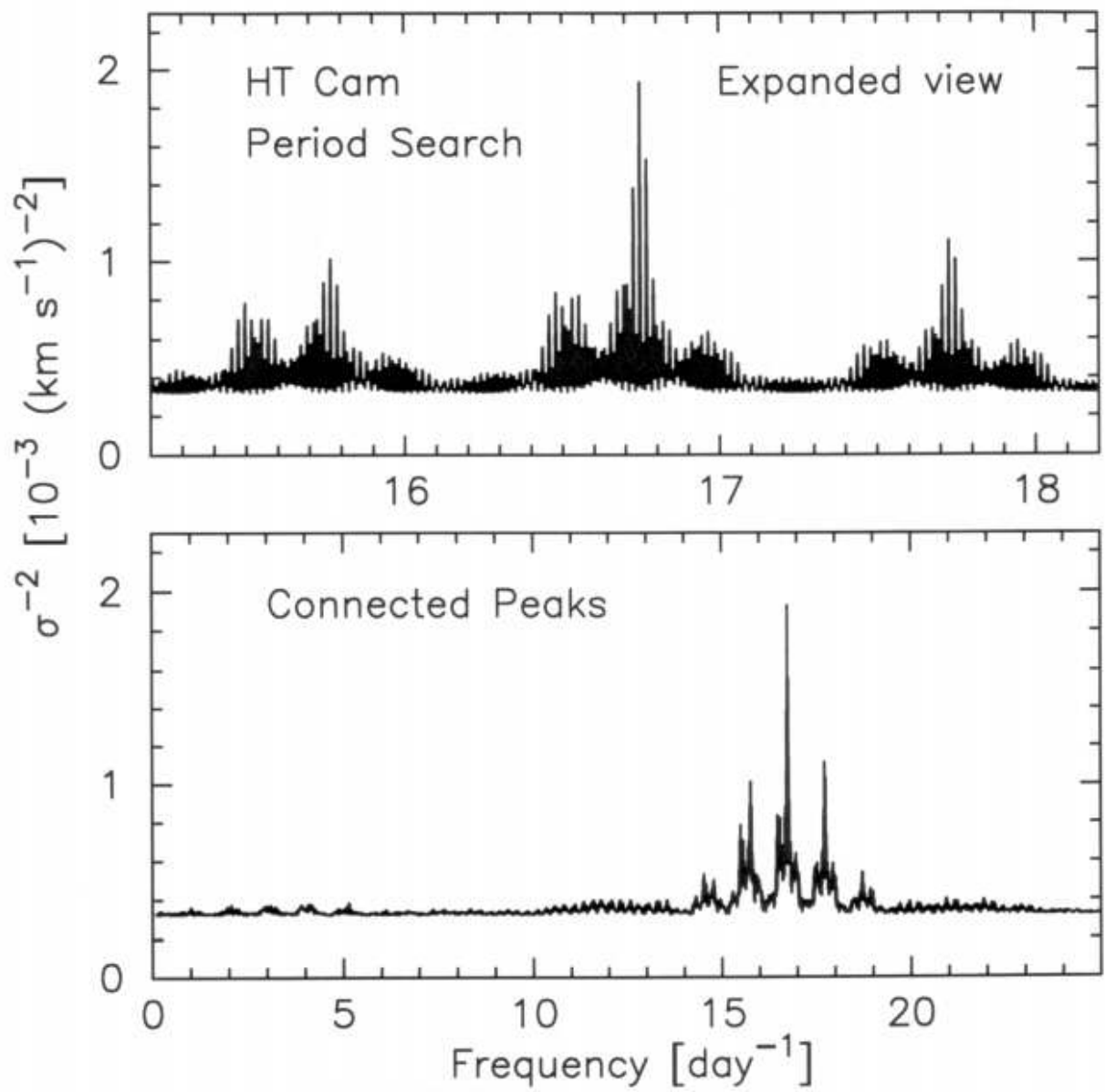

Fig 2

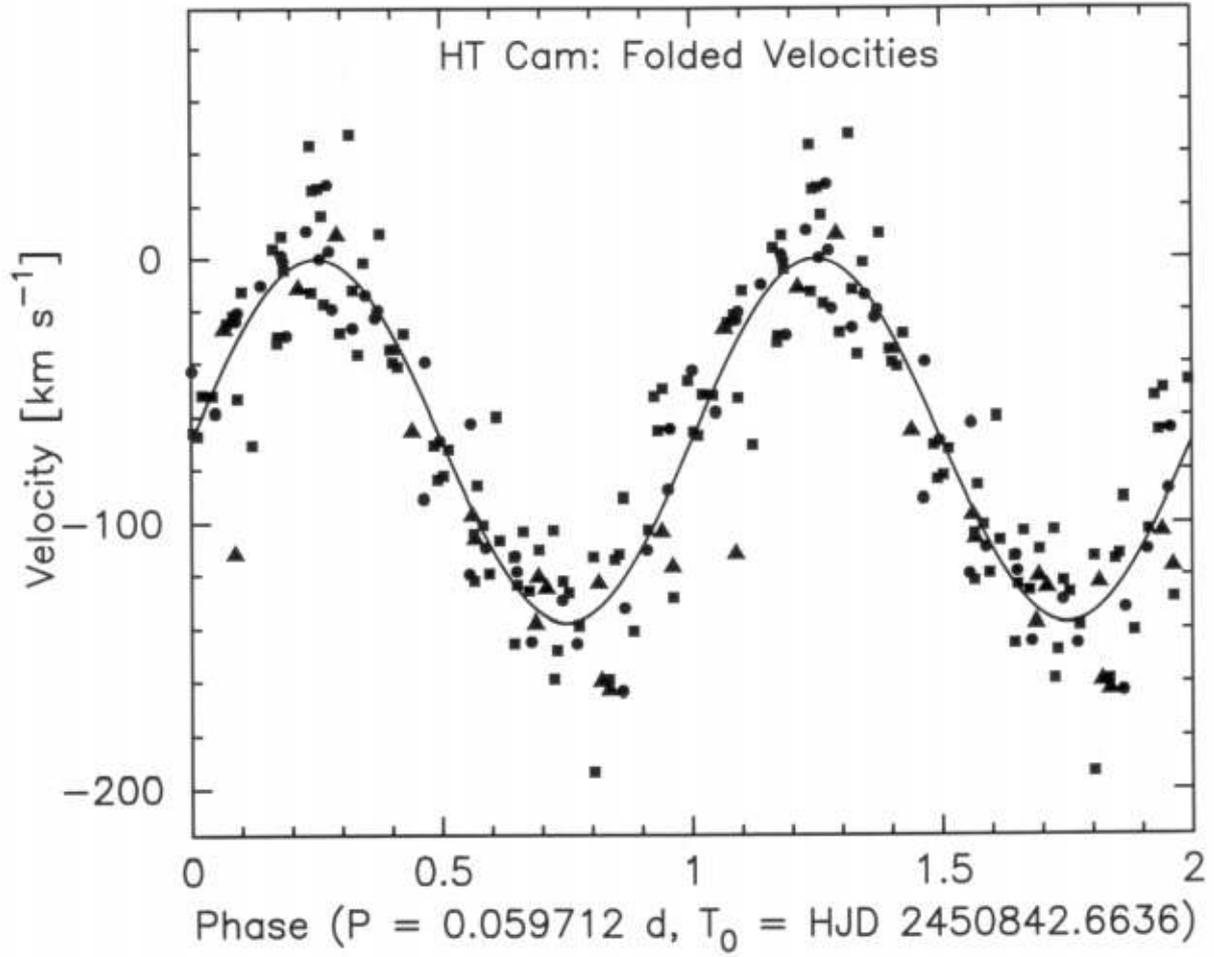

Fig 3

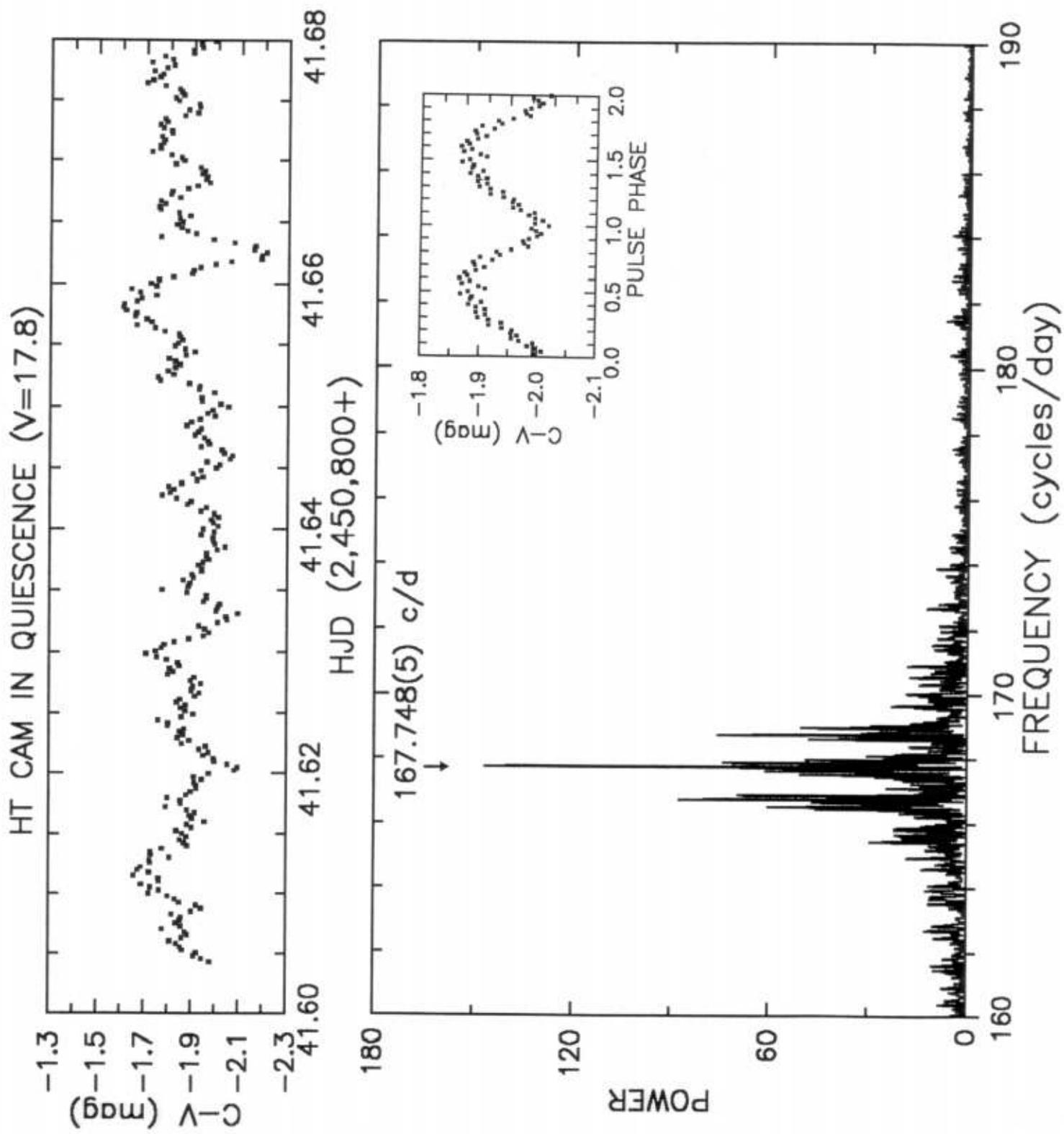

Fig 4

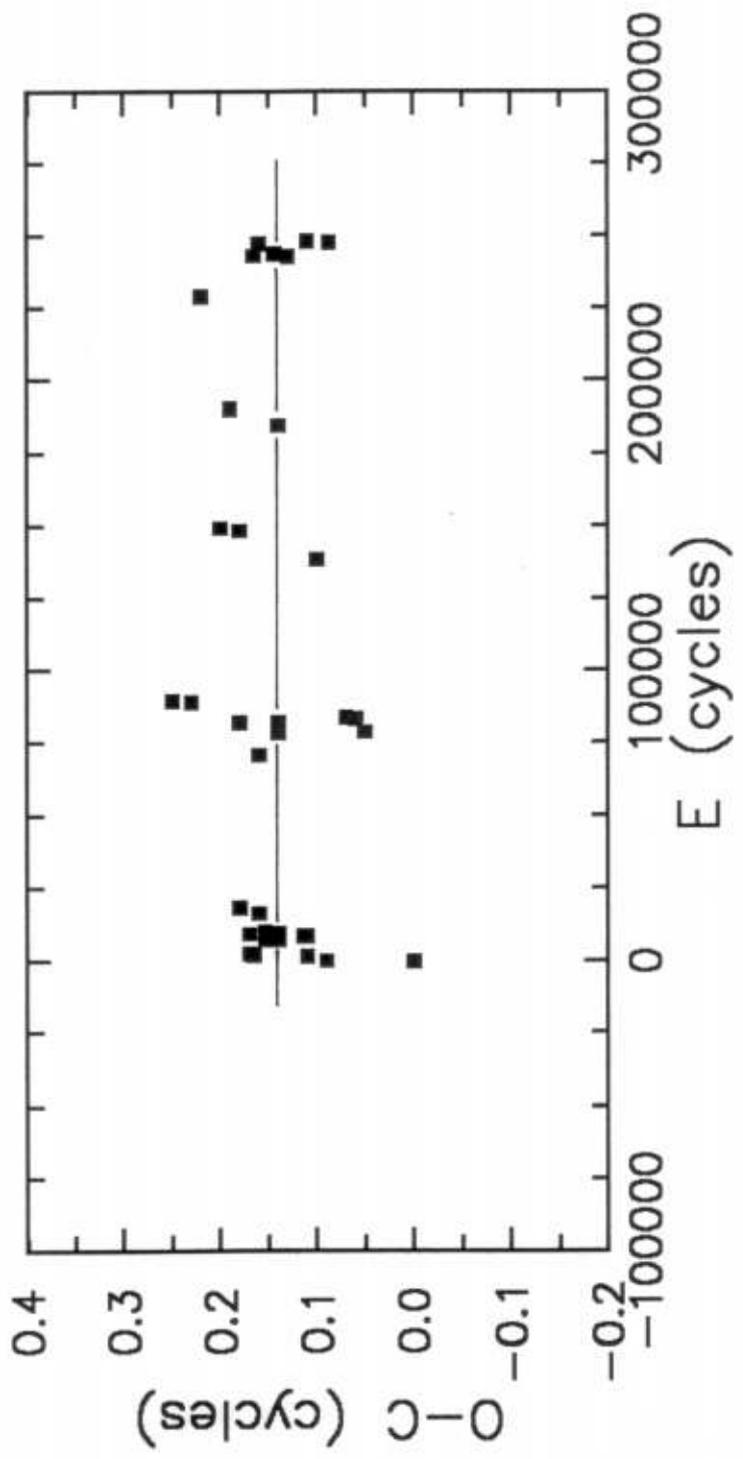

Fig 5

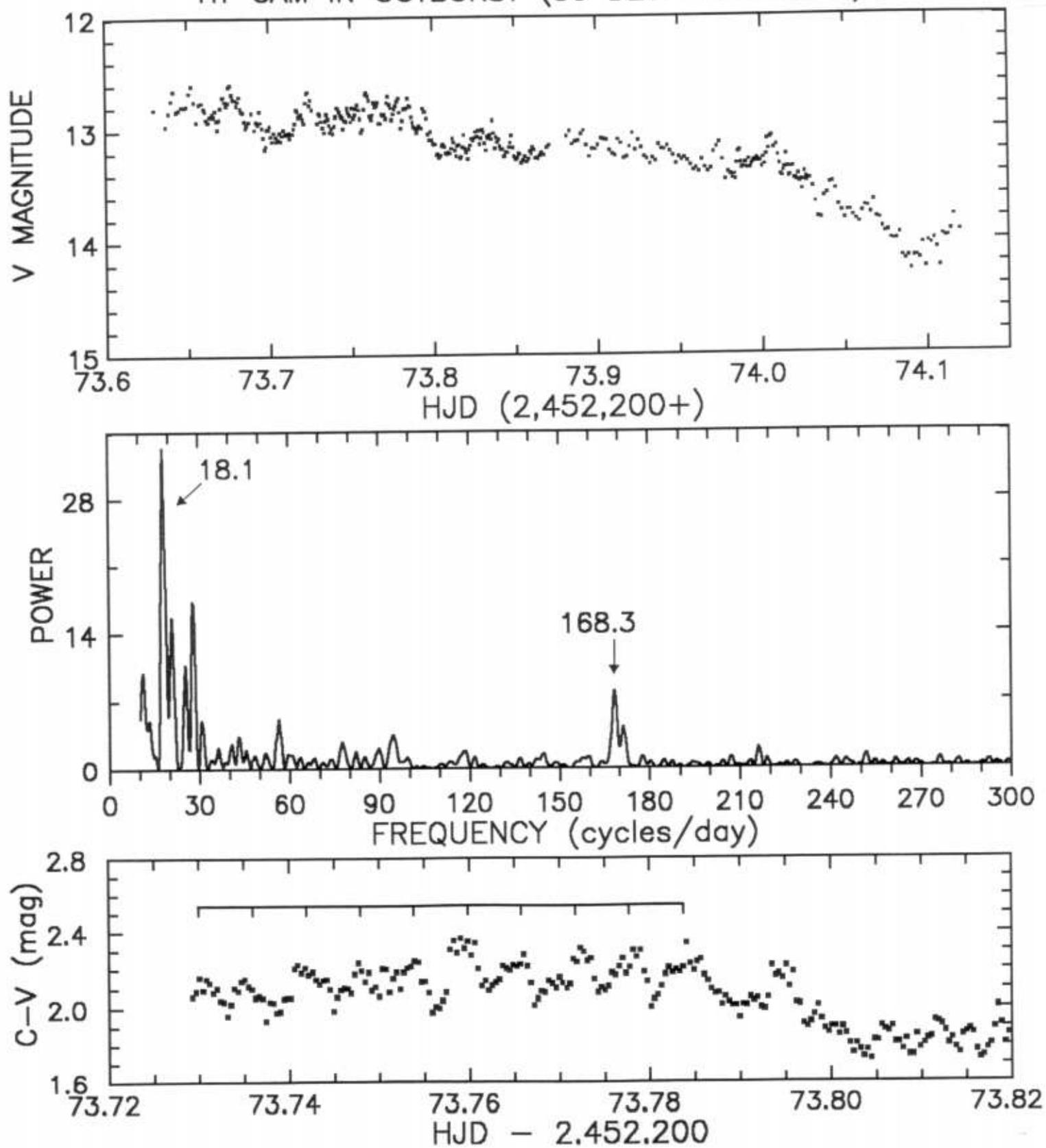

Fig 6